\documentclass[aps,prl,twocolumn,groupedaddress,floatfix,showpacs,showkeys,amsmath,amssymb]{revtex4-1}
\usepackage{graphicx}
\usepackage{bm}
\begin{document}
\title{Plentiful magnetic moments in oxygen deficient SrTiO$_3$}

\author{Alejandro Lopez-Bezanilla$^{1}$}
\email[]{alejandrolb@gmail.com}
\author{P. Ganesh$^{2}$}
\author{Peter B. Littlewood$^{1}$ $^{3}$}
\affiliation{$^{1}$Argonne National Laboratory, 9700 S. Cass Avenue, Lemont, Illinois, 60439, United States}
\affiliation{$^{2}$Center for Nanophase Materials Science, Oak Ridge National Laboratory, One Bethel Valley Road, Tennessee, 37831, United States}
\affiliation{$^{3}$James Franck Institute, University of Chicago, Chicago, Illinois 60637, United States}

\begin{abstract}
Correlated band theory is employed to investigate the magnetic and electronic properties of different arrangements of oxygen di- and tri-vacancy clusters in SrTiO$_3$.
Hole and electron doping of oxygen deficient SrTiO$_3$ yields various degrees of magnetization as a result of the interaction between localized magnetic moments at the defected sites.
Different kinds of Ti atomic orbital hybridization are described as a function of the doping level and defect geometry.
We find that magnetism in SrTiO$_{3-\delta}$ is sensitive to the arrangement of neighbouring vacancy sites, charge carrier density, and  vacancy-vacancy interaction. Permanent magnetic moments in the absence of vacancy doping electrons are observed.
Our description of the charged clusters of oxygen vacancies widens the previous descriptions of mono and multi-vacancies and points out the importance of the controled formation at the atomic level of defects for the realization of transition metal oxide based devices with a desirable magnetic performance.
\end{abstract}

\maketitle
\section{Introduction}

SrTiO$_3$ (STO)  is a wide band gap insulator composed of nominally electrically neutral TiO$_2$ and SrO planes alternately layered in the form of a cubic perovskite.
Alone or in contact with other materials STO has played a prominent role in the unrelenting growing field of oxide superlattices and  transition metal oxide (TMO) electronics\cite{ISI:000188470500037}.
Oxygen vacancies (Ov) are pervasive point-defects in TMOs, and their presence impacts numerous of the material physical properties such as electrical conductivity or magnetism.
Far from representing a drawback in the integration of TMOs in electronic compounds, the atomic control of Ov clusters as well as their density and distribution may lead to purposefully manipulation of the magnetic and electronic properties of a TMO-based material\cite{Crooker}.

Perovskites can accommodate a high density of Ov while keeping pseudocubic structure. Ordered Ov superstructures with a nominal stoichiometry of ABO$_{2.5}$ (brownmillerite) are formed upon removal of a substantial fraction of O atoms. Several experimental studies have revealed the presence of Ov clusters in alkaline-earth titanates, demonstrating that oxygen vacancies in STO can gather in highly n-type doped layers forming ordered domains with enhanced ionic conductivity\cite{Klie2001289}. Atomic precision synthesis of SrTiO$_{3-\delta}$ thin films demonstrated that vacancy-engineered structures can be artificially created. Neutron powder diffraction combined with synchrotron X-ray diffraction techniques have been used to determine the structure of complex-oxide Ti-based structures identifying vacancy ordering arrangements in bulk samples\cite{Suescun}. Switching on/off behavior in TiO$_2$-based resistive-random-access-memories has been found to be driven by the formation and disruption of conducting filaments of Ov\cite{blanka}.

Experimental techniques to increase or deplete charge carrier density Ov-rich areas allow the emergence of magnetic phases in STO based compounds\cite{bi}. With a view to implementing TMO compounds in electronics, a detailed analysis of the properties of the magnetic and conducting states present at the bulk of oxygen-deficient STO is of the uppermost interest. In spite of a considerable literature on oxygen monovacancies, a qualitative systematic description of oxygen multi-vacancies (mOv) in STO under different doping conditions has been lacking.  Although previous studies have dealt with the formation, migration and isolation of Ov, the electronic and magnetic properties of clustered Ov is still far from being fully understood. In particular, the understanding of the Ov geometry and concentration in bulk STO needs to be investigated to correlate the magnetic properties to the atomistic characteristics of the defected material.

In this paper we analyse the large magnetic and electronic tunability of oxygen deficient STO mediated by different doping rates of clustered oxygen vacancies. We use correlated band theory to predict magnetic behaviour in mOv and develop a comprehensive F-center analysis that accounts for the varied defect possibilities in SrTiO$_{3-\delta}$.
Unlike O monovacancies which may exhibit localized magnetic moments for particular doping concentrations or vacancy densities\cite{2014arXiv1408.3103L}, for any doping density we find that isolated mOv always exhibit intrinsic magnetic moments. Both neutral and charged mOv configurations are considered to account for the effect of doping in spin alignment. We find that electrons are confined to the vacated region in between the Ti atoms, producing isolated F-centers whose magnetic moments are very sensitive to both the geometry and the material doping rate.

\section{Computational methodology}
Density-functional-theory based calculations were conducted using the projector augmented-wave method \cite{PhysRevB.50.17953} and the PBE-GGA exchange-correlation functional \cite{PhysRevLett.77.3865}. To improve the description of the electrons occupying the d-orbitals of the Ti atoms at the vacant site, a Hubbard-U correction (GGA+U) as implemented in the VASP code \cite{PhysRevB.48.13115,PhysRevB.54.11169,PhysRevB.59.1758} was included. The rotationally invariant method by Dudarev et al. \cite{PhysRevB.57.1505} with an effective U$_{eff}$=U-J=4.0 eV was applied to capture the strong correlations. The electronic wavefunctions were described using a plane-wave basis set with an energy cutoff of 400 eV. Atomic positions were fully relaxed in large supercells until residual forces were lower than 0.02 eV/\AA. Extra electrons were introduced or removed and compensated with an equally uniform background charge of opposite sign.

\section{Results and discussion}

\subsection{Oxygen mono-vacancy}
We begin by recalling the origin of possible magnetism in the F-center created upon the removal of a single O atom from cubic STO.
When STO is treated at low oxygen partial pressures and high temperatures, Ov are created and balanced by the generation of mobile charges according to the coordination of the oxygen atoms.
In the perovskite structure, O$^{2-}$ anions bond to Sr$^{2+}$ cations at the cube vertex and Ti$^{4+}$ cations at the cube center. In a pristine unit cell a Ti cation is 6-fold coordinated with the anions forming an octahedron. The highest solid state occupied bands contain 18 electrons primarily in oxygen p-orbitals hybridized with both cation orbitals. The lowest unnoccupied bands are composed of Ti 3d t$_{2g}$ states yielding an experimental band gap of 3.2 eV. This picture changes when an O atom is removed from the structure and the crystal symmetry is lowered from cubic $Pm\bar{3}m$ to the C$_{4v}$ point group. In order to minimize the configurational energy the geometry in the vicinity of the vacant site undergoes a strong distortion including a local expansion of the volume of $\sim$1\%, a tilting of the O$_6$ octahedra and an increasing of the Ti-Ti distance upon the formation of a new $\sigma$-bond. Since the O vacates the crystal as a neutral atom the two electrons that accepted from the neighbouring Ti atoms are left in the defected structure, conferring to the Ov a double donor character.

This traditional description of the localized magnetic moments in oxygen-deficient STO is based on density functional theory simulations that considered small supercells and therefore favored the correlation of the defect states. In a previous study\cite{2014arXiv1408.3103L} we reported an extended model to explain the origin and conditions under which magnetic moments sitting at a single Ov are expected to be observed in STO. Therein we demonstrated by means of correlated band theory that magnetism in oxygen-deficient STO is not determined uniquely by the presence of single Ov but by the density of free charge carriers. The local electron correlations between neighbouring oxygen deficient sites, the overlap of their wave functions and the structural rotation of the perovskite octahedra determine the appearence of a magnetic charge distribution confined in the region between two Ti atoms, giving rise to magnetic moments which are further coupled to itinerant conduction band (CB) d-electrons. We found that isolated Ov are nearly hydrogenic double donors, while at higher carrier density, the Ov distorts to accommodate a single bound electron, and therefore a local magnetic moment.

\subsection{Side-by-side oxygen di-vacancy}

\begin{figure}[htp]
 \centering
 \includegraphics[width=0.50 \textwidth]{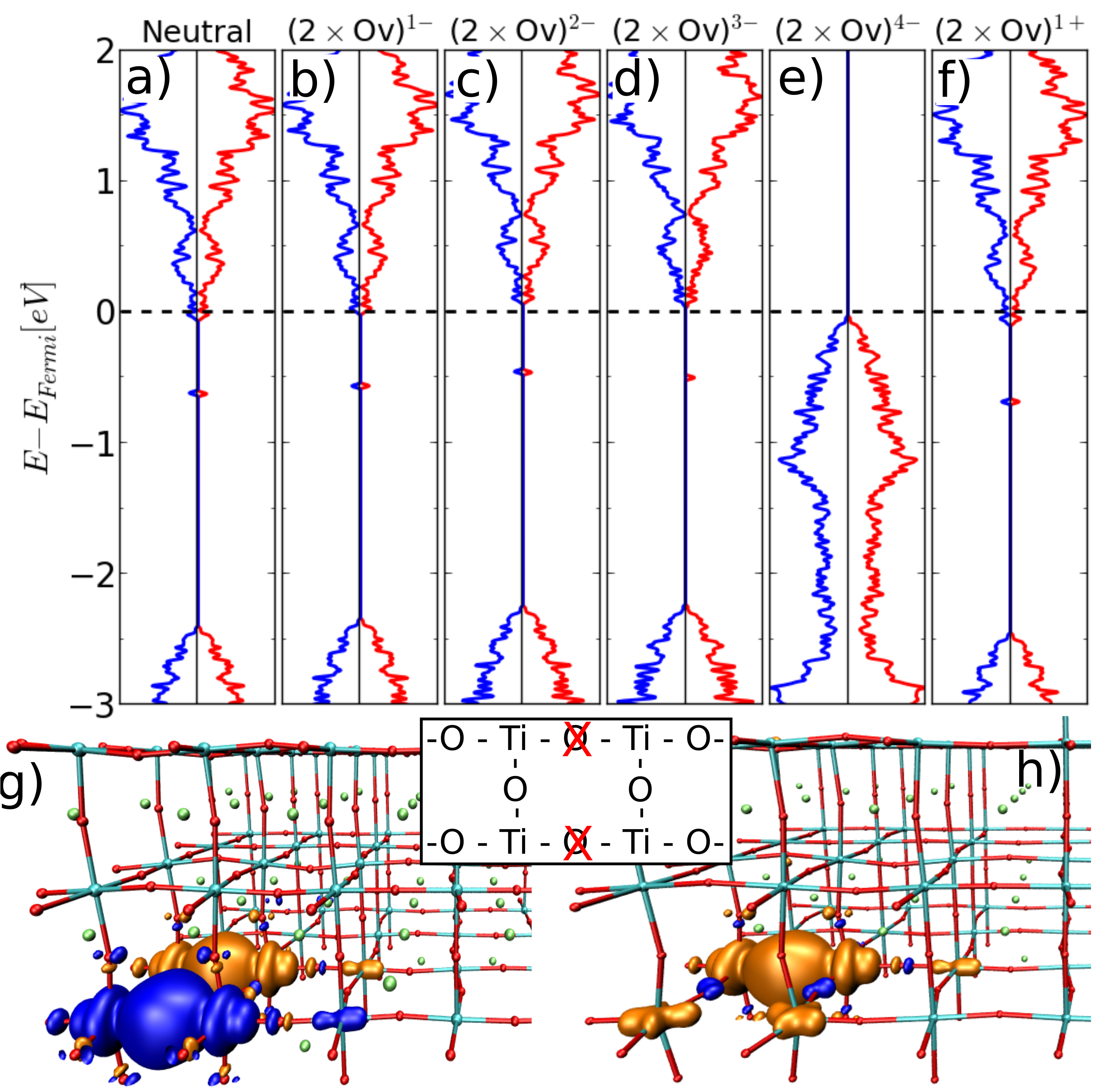}
 \caption{From a) to e), electronic band diagrams displaying the evolution of the electronic states with increasing level of hole doping in a supercell containing two side-by-side oxygen vacancies as schematically shown in the inset. g) The formation of two magnetic states localized at the vacancy sites with opposite spin alignment is robust against doping. h) One electron at a vacant site is preferred over sharing the defect wave function on both site for three-fold hole doping.
 }
 \label{fig1}
\end{figure}

Ov are known to tend to cluster in specific patterns depending on their location either on the bulk or by the slab surface\cite{PhysRevLett.86.4056}. Specific Ov arrangements such as line of Ov are strongly favored and form preferentially along certain crystallographic directions\cite{Ganesh}. This results in a chain of Ti$^{2+}$ cations laterally bonded to only two O atoms. Also a side-by-side ordering where Ti-Ti bonds align in parallel were found to be the most stable magnetic configuration in Ref\cite{Terakura}, where both types of arrangements were found very close in energy. Upon considering several geometries of the supercell containing the Ov clusters we determined that 3$\times$4$\times$7 and 3$\times$3$\times$11 supercells were sufficiently large (depending on the defect geometry) for an accurate description of the different spin alignments at the Ov sites while accounting adequately for the density of states, the octahedra rotation and the electron counting. Due to the reduced lateral extension and the long tail of the F-center wave function along the direction of the vacancy ordering, the longer dimension of the supercell is selected in the Ti-Ti axis. Due the geometry distortion the vacancies may be correlated through the elastic constant of the network, but the large size of the supercell guarantees that the wave functions of periodic images do not overlap either laterally nor in the Ti-Ti bond direction.

To investigate systematically the effects of Ov concentration on the magnetic properties of STO, we start by analysing a di-oxygen vacancy (di-Ov) configuration where two side-by-side in-plane O atoms were removed yielding two parallel Ti-Ti bonds, as depicted schematically in the inset of Figure \ref{fig1}. The removal of two O atoms releases four electrons that are accomodated in the defected bulk similarly as explained above for the monovacancy. One electron is trapped at each of the two vacant sites upon hybridization of the 3d$_{z^2}$ and 4p$_z$ atomic orbitals of each Ti atom, while the remaining electrons delocalize at the CB. This results in two magnetic states with opposite spin alignments and the same energy of $\sim$ -0.5 eV, as exhibited in the band diagram of Figure \ref{fig1}a. The Ti-Ti distance decreases a 1.3\% to 3.91 \AA. A real-space representation of the F-centers localized in between the Ti atoms with different magnetizations are depicted in Figure \ref{fig1}g. Aligning both magnetic moments with same orientation entails an increase of the formation energy of 27 meV.

This picture is barely modified if the system is positively doped by removing one electron. The electron is substracted from the CB, as shown in Figure \ref{fig1}b, and the vacancies remain negatively charged. Additional removal of a charge carrier from the CB leads to a semiconducting system with the magnetic moments intact. This result is similar to the single-hole doped mono-Ov \cite{2014arXiv1408.3103L}, and suggests that up to n-hole doping (n$\equiv$ Ov concentration) the defect levels are pinned at $\sim$ 0.5 eV below the bottom of the conduction band. Further removal of a third electron localizes the remaining electron in a vacant site. It is worth noting that the charge sits solely as an F-center in between a pair of Ti cations separated by 3.93 \AA, whereas the other two Ti atoms move apart from each other a distance of 4.1 \AA\  and slightly contribute to the magnetic moments with a small hybridization of the 3d$_{z^2}$ and 3d$_{xz}$ orbitals (see Figure \ref{fig1}h). If both vancant sites are forced to share the single electron wave function, the single state observed in Figure \ref{fig1}d splits yielding two states with equal energy and different spin if the sites couple antiferromagneticly, or with different energy and same spin alignment if coupled ferromagneticaly. In both cases the formation energy of the defected system increases 200 meV. Finally, removing 4 electrons the oxygen deficient structure becomes a wide gap insulator as bulk STO. On the contrary, negative doping injects charge carriers in the bottom of the Ti d-bands and consequently enhances the metallic character by shifting up the Fermi level higher in the CB.

\subsection{Head-to-head oxygen di-vacancy}

\begin{figure}[htp]
 \centering
 \includegraphics[width=0.45 \textwidth]{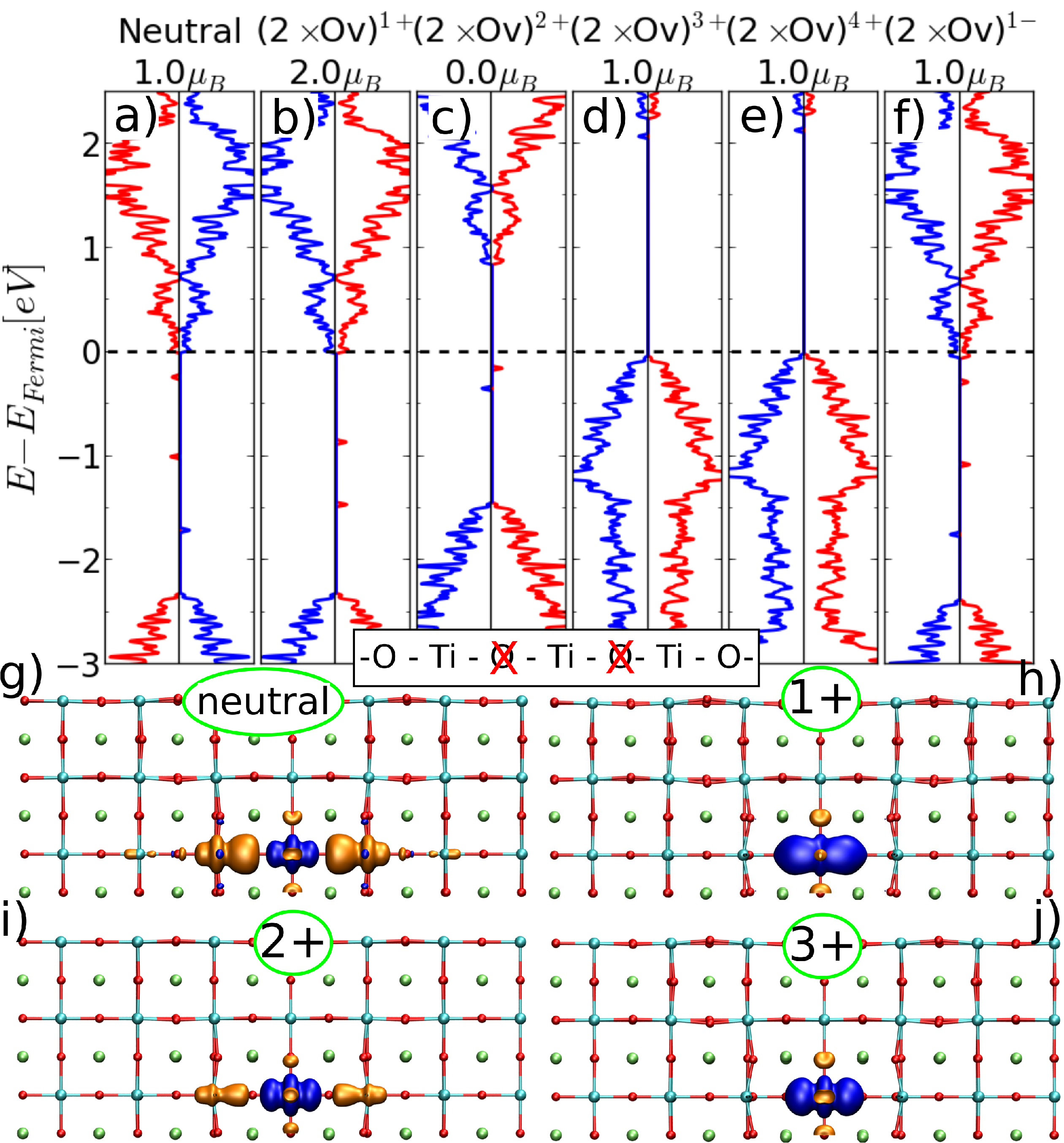}
 \caption{From a) to e), electronic band diagrams showing the evolution of the electronic states with increasing level of hole doping of a supercell containing two head-to-head oxygen vacancies. f) corresponds to a negatively doped configuration. The metallic or insulating behaviour, the total magnetic moment, and the distribution of the magnetic states strongly depend on the doping level. The real-space representation of the net charge density of the undoped (neutral) configuration is displayed in g), for singly hole doped in h), for two-fold hole doped in i), and for three- and four-fold hole doped in j). The latter also represents a hybrid 4s3d$_{z^2}$ state for a di-oxygen vacancy in a supercell with four electrons removed. The excess of spin-up with respect to spin-down charge density in a  supercell is plotted in g), h) and j) for an isosurface value of 1.7 $\times$10$^{-3}$e$^-$/\AA$^3$, and in i) of 1.7 $\times$10$^{-4}$e$^-$/\AA$^3$.}
 \label{fig2}
\end{figure}

An alternative arrangement of two Ov is a head-to-head configuration, which causes the reduction of two Ti$^{4+}$ cations to two Ti$^{3+}$ cations bridged by a Ti$^{2+}$ cation: -Ti$^{3+}$-Ov-Ti$^{2+}$-Ov-Ti$^{3+}$- (see inset of Figure \ref{fig2}). This arrangement is very close in energy to the side-by-side configuration discussed above and the computational approach becomes decisive in determining the real ground state\cite{Terakura}. Despite all that, Ov in the proximity of domain wall and grain boundaries have been identified as energetically preferred when segregated in linear clusters\cite{PhysRevLett.86.4056}. Starting from the ideal structure of STO the creation of this type di-Ov maintains the original framework of the perovskite structure although some octahedra tiltings around the defective sites are observed.

\begin{figure}[htp]
 \centering
 \includegraphics[width=0.45 \textwidth]{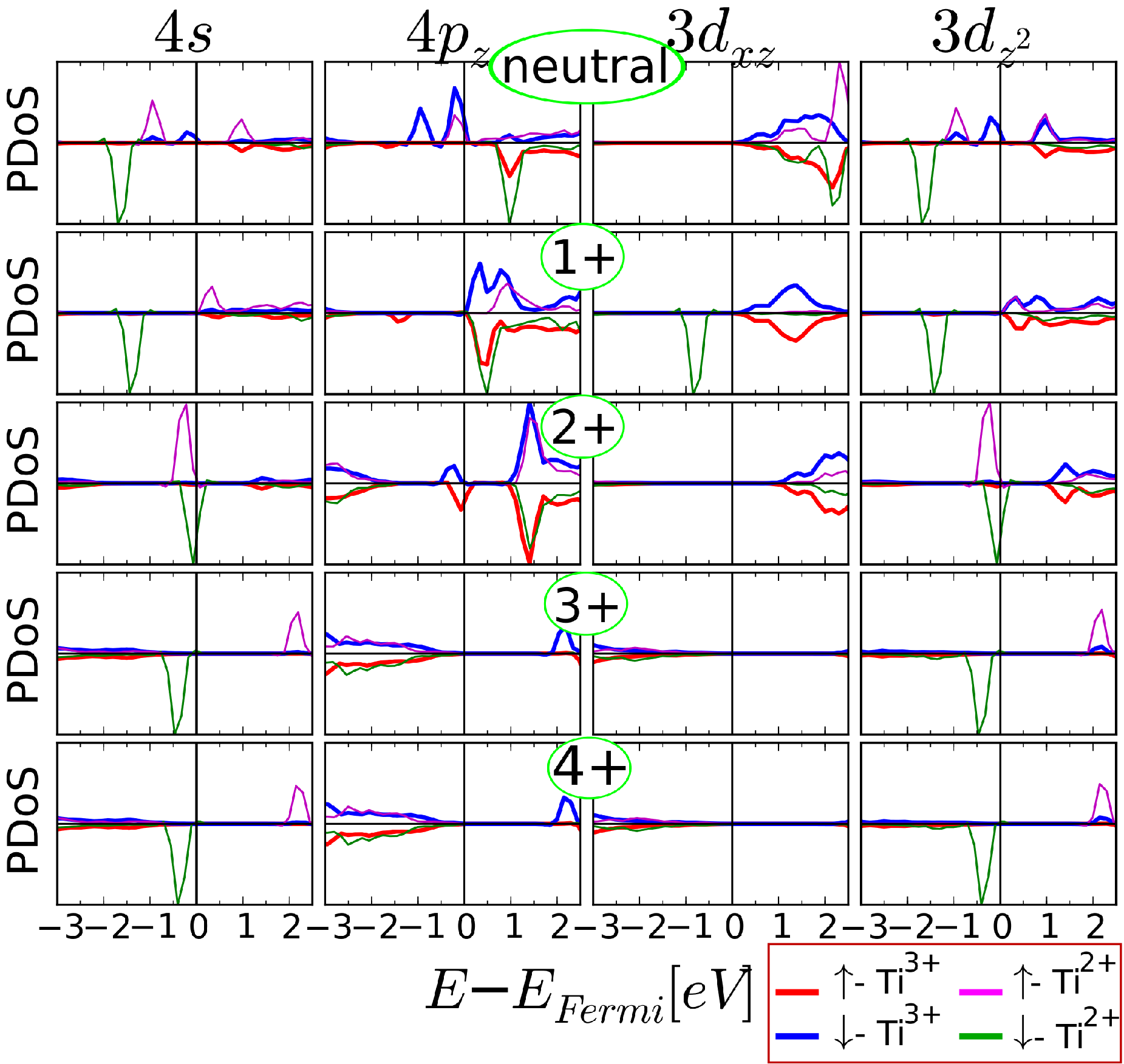}
 \caption{Spin-resolved projected density of states (PDoS) on the  4s, 4p$_z$, 3d$_{xz}$ and 3d$_{z^2}$ orbitals of the Ti$^{3+}$ and Ti$^{2+}$ cations at the defect shown in Figure \ref{fig4}, consisting in two oxygen vacancies aligned in a SrTiO$_3$ slab. From upper to lower panels, the PDoS correspond to the neutral, single, two-, three-, and four-fold positively doped configurations respectively. }
 \label{fig3}
\end{figure}

According to the symmetry and the oxidation rate, different types of atomic hybridization can be observed. The band diagram of Figure \ref{fig2}a shows that three electrons are trapped in three in-gap states whereas the fourth electron is delocalized in the CB, rendering the system metallic. The projected density of states (PDoS) onto the Ti atoms is employed to assess the degree of electron localization on each atom, giving the the contribution of each cation orbital to the total density of states at a particular energy. Figure \ref{fig3} shows that the magnetic state at $\sim$-1.7 eV is a hybrid of the 4s and 3d$_{z^2}$ orbitals of uniquely the Ti$^{2+}$ cation. The same two orbitals along with the p$_z$ and the 3d$_{z^2}$ orbitals of the Ti$^{3+}$ cations define the localized state at $\sim$-1.5 eV, which has opposite spin to the latter. Finally, the shallow state at $\sim$-0.2 eV is the result of mixing the p$_z$ orbitals of both type of cations and the 3d$_{z^2}$ orbital of the Ti$^{3+}$ cations. A real-space representation of the net charge occupying the space along the Ti-Ov-Ti-Ov-Ti axis is given in Figure \ref{fig2}g. The total magnetic moment of the defect is 1 $\mu_B$. The distance between the Ti$^{2+}$ and a Ti$^{3+}$ cation is of 3.90\AA, and from a Ti$^{3+}$ to the Ti$^{4+}$ cation along the defect axis is of 4.00 \AA.

The removal of an electron out of the 2676 contained in the supercell has extraordinary consequences on both the local chemistry and the magnetism of the di-Ov. Indeed, unlike the oxygen monovacancies and the di-Ov presented above, instead of a CB electron a localized state is removed when an electron is extracted from the supercell as observed in Figure \ref{fig2}b. Interestingly, instead of reducing the magnetic moment removing an electron enhances the total magnetic moment up to 2 $\mu_B$. The explanation of such a drastic change relies on the hybridization undergone by the Ti cations' orbitals at the defect sites. As shown by the PDoS in Figure \ref{fig3}, the Ti$^{3+}$ cations has no contribution to the two in-gap states and, in turns, the central Ti cation orbitals accomodate two electrons. The deeper in-gap state at $\sim$-1.5 eV is formed by a hybridization of the 4s and 3d$_{z^2}$-orbitals, whereas the orbital 3d$_{xz}$ accomodates by itself alone an electron at $\sim$-0.9 eV. Notice that the molecular-like orbital formed upon the combination of the three cations vanishes and only one cation holds the magnetic moment of the entire system (Figure \ref{fig3}h). The distance between the central cation holding the excess of charge and the neighbouring Ti$^{3+}$ cations along the defect line increases to 4.01 \AA, while the Ti$^{3+}$-Ti$^{4+}$ distance is reduced to 3.93 \AA.

Further removal of an electron renders the system a non-magnetic insulator. The charge is withdrawn from the CB and a magnetic state with similar spatial extension to the neutral configuration state (Figure \ref{fig2}i) is formed by the concurrent addition of the 4s, 4p$_z$, and 3d$_{z^2}$-orbitals of both the Ti$^{3+}$ and Ti$^{2+}$ cations. As a result, two magnetic states of opposite spin sign lay in the middle of the band gap. Observing the Figure \ref{fig2}d one might conclude that an additional extraction of an electron from the supercell removes all localized in-gap states. But a detailed inspection of the corresponding PDoS diagram shows that a flat state in resonance with the valence band (VB) states and formed upon hybridization of the 4s and 3d$_{z^2}$ central Ti cation orbitals is responsible of the supercell magnetic moment of 1 $\mu_B$. At high energy, close to the CB, the antibonding counterpart is observed. This spatially localized state is plotted in Figure \ref{fig2}j, where the geometry is barely modified with respect to the previous singly doped structure.

Interestingly, this is the same state that is observed when a total of four electrons are removed from the system. Contrary to what observed in the side-by-side di-Ov, the removal of all the Ov doping electrons does not  necessarily entails the absence of a localized state and a magnetic moment, but the formation of a permanent magnetic moment sitting on the central Ti$^{2+}$ cation. Finally, negative doping of the supercell with an additional electron shifts the Fermi level enhancing the metalic character of the oxygen deficient system, as shown in Figure \ref{fig2}f.

\subsection{Head-to-head oxygen tri-vacancy}

\begin{figure}[htp]
 \centering
 \includegraphics[width=0.45 \textwidth]{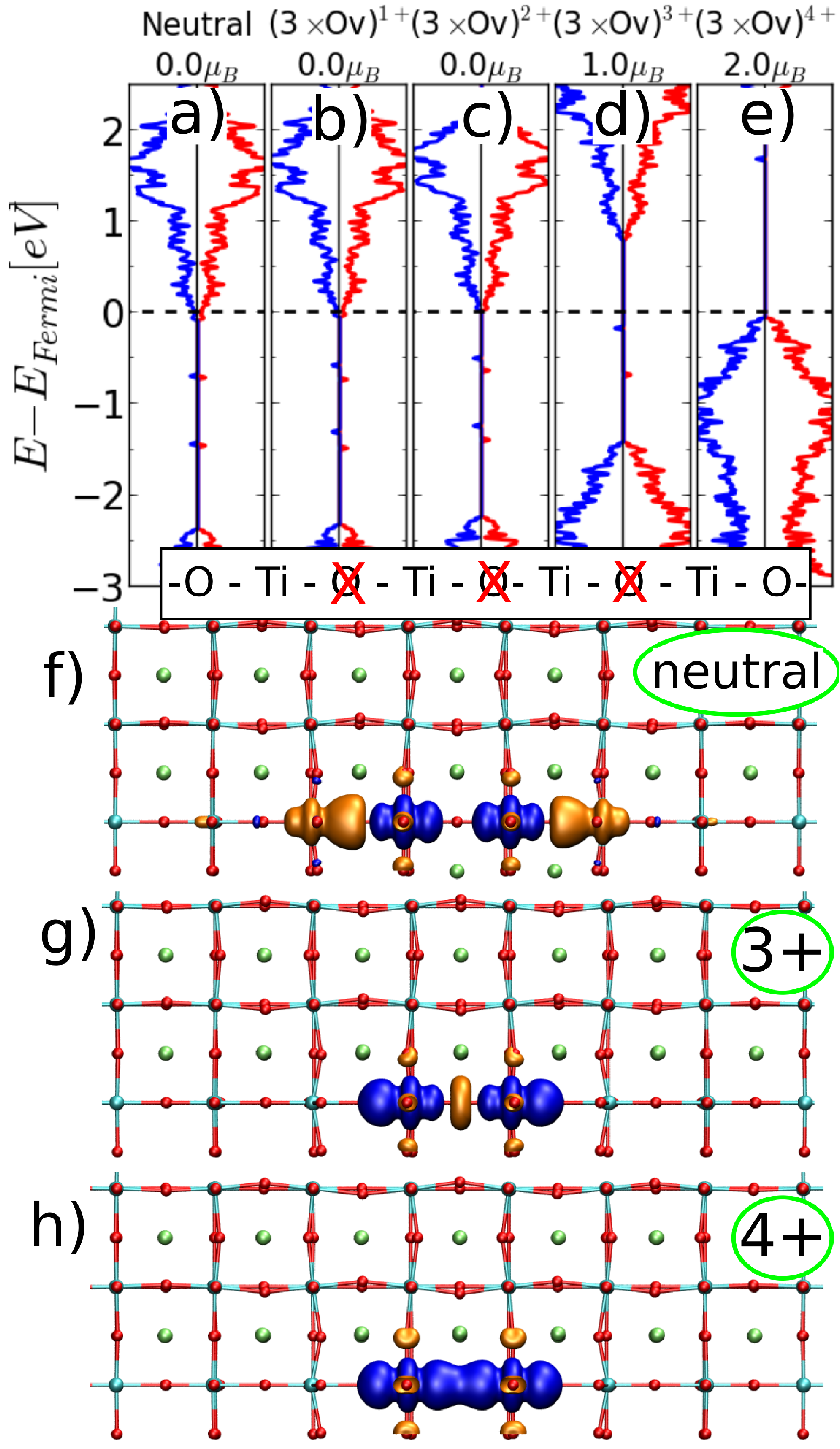}
 \caption{From a) to e), electronic band diagrams showing the evolution of the electronic states with increasing level of positive doping of a supercell containing three head-to-head oxygen vacancies. The metallic or insulating behaviour, the total magnetic moment, and the distribution of the magnetic states strongly depend on the doping level. Panels from g) to j) display a real-space representation of the net charge density of the SrTiO$_{3-\delta}$ configurations. Real-space representations of the net charge density of the SrTiO$_{3-\delta}$ configurations are in display in f) for neutral, single, and two-fold hole doped defected supercell, and in g) and h) for three- and four-fold hole doped configuration respectively. The excess of spin-up with respect to spin-down charge density in a  supercell is plotted in f) for an isosurface value of 2.6 $\times$10$^{-5}$e$^-$/\AA$^3$, and in g) and h) of 1.6 $\times$10$^{-3}$e$^-$/\AA$^3$.}
 \label{fig4}
\end{figure}

\begin{figure}[htp]
 \centering
 \includegraphics[width=0.45 \textwidth]{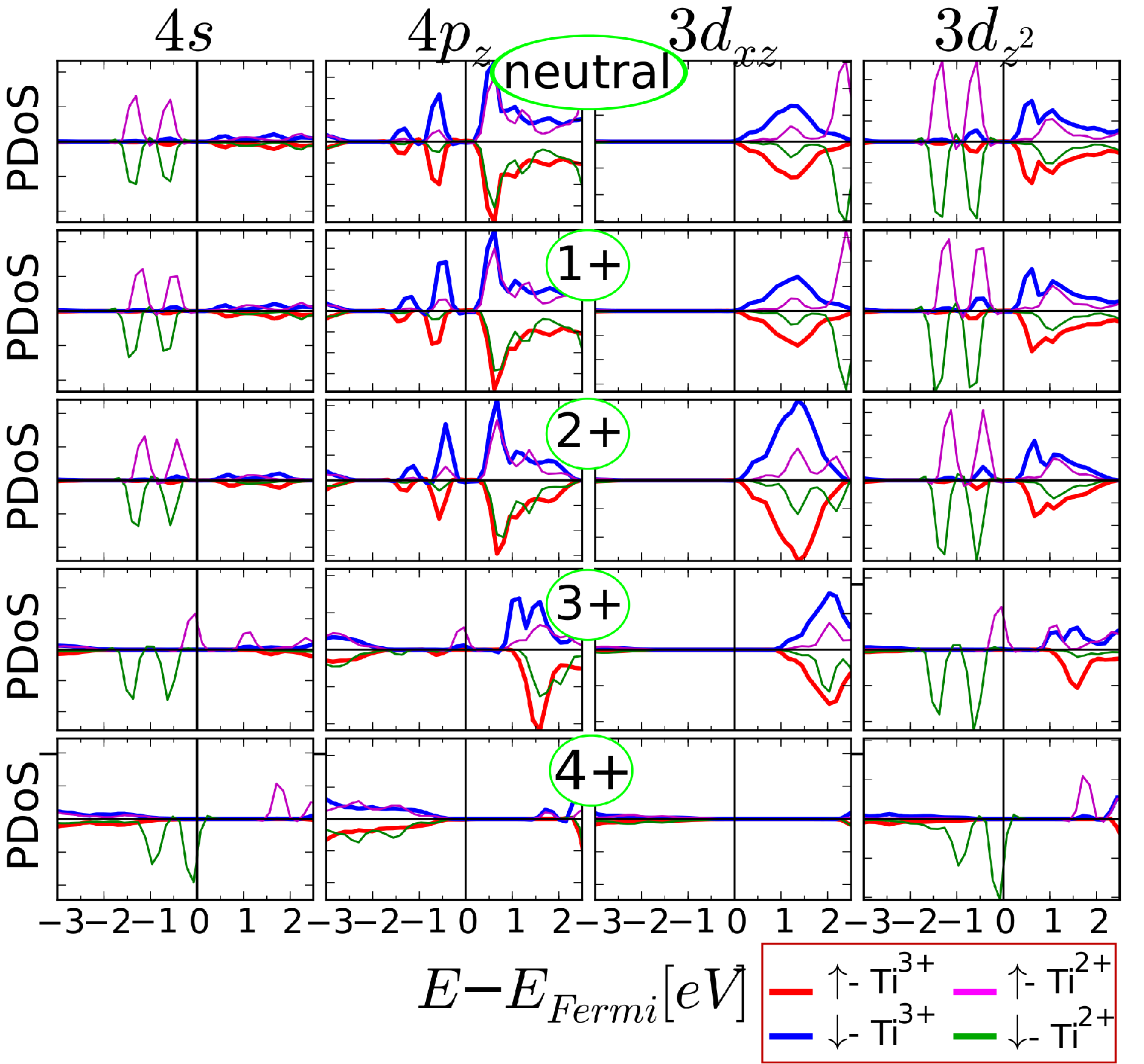}
 \caption{Spin-resolved projected density of states (PDoS) on the  4s, 4p$_z$, 3d$_{xz}$ and 3d$_{z^2}$ orbitals of the Ti$^{3+}$ and Ti$^{2+}$ cations at the defect shown in Figure \ref{fig4}, consisting in three oxygen vacancies aligned in a SrTiO$_3$ slab. From upper to lower panels, the PDoS correspond to the neutral, single, two-, three-, and four-fold positively doped configurations respectively.}
 \label{fig5}
\end{figure}

Next, we follow the same procedure to analyse the changes in the magnetic moment of three aligned Ov upon doping the supercell. Removing three O atoms as depicted in the inset of Figure \ref{fig4}, the supercell is left with two Ti$^{3+}$ and two Ti$^{2+}$ cations. This type of defect exhibits a strong doping character due to the six electrons that are released. Figure \ref{fig4}a shows that the neutral configuration corresponds to a metallic system with two electrons in the CB and two groups of flat bands each containing two in-gap states energetically degenerate and with opposite spins. Thereby the supercell exhibits a zero overall magnetization with several magnetic moments localized in the Ti cations. Figure \ref{fig4}f displays the Ti$^{2+}$ cations holding electronic charge with opposite spin to the Ti$^{3+}$ cations, and Figure \ref{fig5} the substantial differences in the contribution of the cation orbitals to the localized states. The PDoS reveal that the four in-gap states are formed by the combination of the 4s, 4p$_z$, and 3d$_{z^2}$-orbitals of both types of cations. The weight of the Ti$^{3+}$ p$_z$-orbitals in the hybridization is larger than those of the Ti$^{2+}$ cations, and conversely for the Ti$^{2+}$ 3d$_{z^2}$-orbitals. The degeneracy in energy of the magnetic moments is lifted when the Fermi energy shifts down when one electron is removed from the conduction band (Figure \ref{fig4}b). The defected system becomes insulating when a second electron is removed. Note that in no case the distribution of the net charge density is modified by the hole doping and both configurations exhibit the same magnetic moments as the neutral configuration, and an overall zero magnetic moment. Also the distances between cations along the defected line are conserved through the three doping strengths: The distance between the Ti$^{2+}$ cations increases to 4.30\AA, from a Ti$^{2+}$ to a Ti$^{3+}$ cation reduces to 3.72 \AA, and from a Ti$^{3+}$ to a Ti$^{4+}$ cation increases to 4.00 \AA, with respect to bulk structure.

This type of defected STO becomes insulating upon removal of three electrons. The remaing three electrons are accomodated in the same number of in-gap localized states, as shown in the band diagram of Figure \ref{fig4}d, and the corresponding PDoS of Figure \ref{fig5}. The localized state with higher energy is composed of 4s, 4p$_z$, and 3d$_{z^2}$-orbitals of the  Ti$^{2+}$ cation, whereas the state with opposite spin sign at $\sim$ -0.5 eV below the latter is a combination of the 4s and 3d$_{z^2}$-orbitals of the same type of cation. A state deeper in energy and composed of the same orbitals and spin sign than the later is found in resonance with the VB states. The resulting net charge distribution in the real space displayed in Figure \ref{fig4}g shows a charge density with opposite spin to the localized moments that it separates which, in turns, vanishes when a forth electron is removed from the supercell. Indeed, the former state becomes empty with an energy close to the CB states (see Figure \ref{fig4}e) leaving the defected area with net charge density joining both Ti$^{2+}$ cations with no nodes, as plotted in Figure \ref{fig4}h. Note that one of the consequences of depleting electrons from the defected structure is an increasing on the total magnetic moment, which evolves from zero up to 2 $\mu_B$ in the four-fold hole-doped configuration.

\section{Conclusion}

In spite of the broad interest in STO as a model perovskite material, a detailed description of the magnetic properties of oxygen-defected STO in bulk STO was lacking. We have provided a systematic study of the magnetism arising under different doping conditions of multiple isolated Ov clustered in various geometries to show that the local chemistry of the defect is strongly dependent on the defected material doping level. Localized magnetic moments may exhibit a large tunability upon modification of the charge carriers concentration. Unlike mono-Ov and depending on the mOv geometry, permanent magnetic moments can be induced even in the absence of intrinsic doping. These results are consistent with recent experiments on complex oxides heterostructures where the  magnetism in STO is strongly dependent of the charge carrier concentration with the observation of ferromagnetism in electron depleted regions\cite{bi}.
The understanding and control of the various doping levels of Ov and their relative proximity is clearly critical to evaluating the possible role of these centers in the magnetic behavior of STO under different doping conditions.
Correlation effects between defected sites can lead to new functionalities in designed TMO-based structures.
The implications of our results may be generalized to other defective perovskite-based systems that may now be expected to exhibit a similar behavior.

\section{Acknowledgments}

\begin{acknowledgments}
We acknowledge the computing resources provided on Blues high-performance computing cluster operated by the Laboratory Computing Resource Center at Argonne National Laboratory. Work at Argonne is supported by DOE-BES under Contract No. DE-AC02-06CH11357. PG was sponsored by the Laboratory Directed Research and Development Program of Oak Ridge National Laboratory, managed by UT-Battelle,LLC,for the US Department of Energy. J. Low's technical support is gratefully acknowledged.
\end{acknowledgments}

\end{document}